\documentclass[sigconf]{acmart}

\renewcommand\footnotetextcopyrightpermission[1]{} 
\usepackage{cite}
\usepackage{graphicx}
\usepackage{float}
\usepackage{subfigure}
\usepackage{url}

\begin{document}

\title{CIR at the NTCIR-17 ULTRE-2 Task}

\author{Lulu Yu}
\affiliation{
 \institution{CAS Key Lab of Network Data \\ Science and Technology, ICT, CAS}
 \institution{University of Chinese Academy of Sciences}
 \city{Beijing}
 \country{China}
}
\email{yululu23s@ict.ac.cn}

\author{Keping Bi}
\affiliation{
 \institution{CAS Key Lab of Network Data \\ Science and Technology, ICT, CAS}
 \institution{University of Chinese Academy of Sciences}
 \city{Beijing}
 \country{China}
}
\email{bikeping@ict.ac.cn}

\author{Jiafeng Guo}
\affiliation{
 \institution{CAS Key Lab of Network Data \\ Science and Technology, ICT, CAS}
 \institution{University of Chinese Academy of Sciences}
 \city{Beijing}
 \country{China}
}
\email{guojiafeng@ict.ac.cn}

\author{Xueqi Cheng}
\affiliation{
 \institution{CAS Key Lab of Network Data \\ Science and Technology, ICT, CAS}
 \institution{University of Chinese Academy of Sciences}
 \city{Beijing}
 \country{China}
}
\email{cxq@ict.ac.cn}
\begin{abstract}
The \textbf{C}hinese academy of sciences \textbf{I}nformation \textbf{R}etrieval team (CIR) has participated in the NTCIR-17 ULTRE-2 task. This paper describes our approaches and reports our results on the ULTRE-2 task. We recognize the issue of false negatives in the Baidu search data in this competition is very severe, much more severe than position bias. Hence, we adopt the Dual Learning Algorithm (DLA) to address the position bias and use it as an auxiliary model to study how to alleviate the false negative issue. We approach the problem from two perspectives: 1) correcting the labels for non-clicked items by a relevance judgment model trained from DLA, and learn a new ranker that is initialized from DLA; 2) including random documents as true negatives and documents that have partial matching as hard negatives. Both methods can enhance the model performance and our best method has achieved nDCG@10 of $0.5355$, which is 2.66\% better than the best score from the organizer.
\end{abstract}
\maketitle
\pagestyle{plain} 
\section*{Team Name}
CIR
\section*{Subtasks}
Effectiveness Evaluation Subtask (Chinese), Robustness Investigation Subtask (Chinese)

\section{Introduction}
Learning to Rank (LTR) is a crucial component in various real-world systems, especially search engines. While human annotations provide accurate relevance information, they are expensive to obtain. User clicks in the search logs can serve as an indication of document relevance and are cheap to collect. This makes them favorable for modern ranking models, which typically have numerous parameters, and require a substantial amount of training data. However, it is obvious that severe bias and noise exist in the click data e.g. position bias. Thus, extensive research on unbiased learning to rank (ULTR) \citep{wang2016learning, ai2018unbiased, ai2021unbiased} has been conducted to mitigate different biases.

There is a lack of public datasets that have various types of bias from real-world user interactions, so existing ULTR methods have been primarily studied on synthetic datasets and achieved promising results. For instance, Joachims et al. \citep{joachims2017unbiased} introduced a randomization experiment to estimate position bias and used Inverse Propensity Weighting (IPW) to debias click signals. To eliminate the need for result randomization, Ai et al. \citep{ai2018unbiased} proposed the Dual Learning Algorithm (DLA) to jointly learn an unbiased ranker and an unbiased propensity model. However, it remains uncertain whether these methods would be effective when applied to real-world datasets. Given this observation, the ULTRE-2 task encourages participants to explore ULTR approaches to alleviate various types of biases in real user clicks during training, and achieve better ranking performance based on the search dataset collected from the search logs from the largest Chinese search engine, Baidu.

The biases in real-world search datasets are much more complex than the position bias in the simulated clicked data according to the position model. Therefore, the existing popular ULTR methods that handle position bias only do not work effectively on the Baidu search dataset as expected \citep{yu2023feature}. We attribute this to that UTLR methods mostly address the position bias on the clicked documents and treat non-clicked items as irrelevant while most non-clicked items in the first search result page (SERP) are relevant. The items are not clicked because they are not examined rather than not being relevant. Search engines typically provide high-quality first SERPs where most items are relevant to the query, especially for head queries whose click data is abundant. The false negative issues can be much more severe than the position bias in real-world search data. Hence, Instead of focusing on the position bias, in our solution, we mainly focus on how to address the false negative issue. We simply adopt the state-of-the-art ULTR method DLA \citep{ai2018unbiased} to alleviate position bias as it has shown superior performance compared to other methods \citep{zou2022large}, and use it as the foundation model to address the issue of false negatives.

We probe the issue from two perspectives. The first is selection bias that represents the bias towards the documents that have zero examination propensities. Usually, selection bias \citep{cai2022hard} is used to describe the case that some documents are not selected by the ranker on the result page and thus have no chance to be examined. In this task, documents at lower positions have similar issues, so we approach the issue from the perspective of selection bias and aim to correct the labels of the non-clicked items. The second one is negative sampling. ULTR models are usually trained only using the results on the SERPs \citep{ai2018unbiased, wang2016learning}. However, we know that there are many false negatives in this data which would confuse the model. Introducing true negatives for model training can guide the model to learn to differentiate relevance. 

Concretely, for selection bias, we employ the model trained with DLA as the relevance judgment model to generate labels for non-clicked items. Then, based on the updated labels, we train another model that is initiated from the judgment model to learn true relevance. We refer to this approach as a Dual Learning Algorithm with Label Correction (DLA-LC). For negative sampling, we sample random documents as true irrelevant documents and also involve other documents that are partially relevant to act as hard negatives. Our experimental results show that both approaches can boost the model performance.
\begin{table}[htbp]
  \caption{A summary of the notations used in this paper. }
  \label{tab:notations}
  \begin{tabular}{cl}
    \toprule
    Notation & Meaning\\
    \midrule
    $q$ & one query\\
    $d$ & one candidate item \\
    $\pi_q$ & the candidate document list of query $q$ \\
    $o,r,c$ & \begin{tabular}[c]{@{}l@{}}
    binary random variables indicating whether \\
    an item d is observed, perceived as relevant, \\
    and clicked by a user
    \end{tabular}   \\
  \bottomrule
\end{tabular}
\end{table}
\section{RELATED WORK}
To leverage biased user clicks for optimizing an unbiased ranking model in learning-to-rank systems, considerable research has been tailored for unbiased learning to rank (ULTR) to alleviate the bias present in user clicks. There are two streams of ULTR methodologies. One trend depends on click modeling \citep{chuklin2022click, craswell2008experimental}, which makes assumptions about user browsing behaviors and then models the examination probability. By maximizing the likelihood of the observed data, they can accurately infer relevance from user clicks. For example, Craswell et al. \citep{craswell2008experimental} designed a Position-Based Model that assumes users will click on a document if and only if they have examined it and considered it relevant. However, this method is based on statistics requiring multiple occurrences of the same query-document pair for reliable inference, which is challenging for long-tail queries and sparse query systems e.g. personal search. The other trend derives from counterfactual learning, which treats bias as a counterfactual factor and debiases user clicks via inverse propensity weighting \citep{ai2021unbiased}. The latest work is that Ai et al. \citep{ai2018unbiased} presents a Dual Learning Algorithm that treats training an unbiased ranking model and propensity model as a dual problem and optimizes each other together. Most ULTR approaches experiment on synthetic datasets and perform effectively, but whether they can maintain the performance on real user clicks still remains uncertain. In this competition, we propose a DLA-LC to make DLA more suitable for real-world datasets.
\section{UNBIASED LEARNING}
In this section, we first introduce our methods to address the position bias and false negative issues and we also describe one upper bound model training we use.
\subsection{Alleviating Position Bias with IPW}
To alleviate the position bias, a series of IPW-based methods \citep{ai2021unbiased} have been proposed and shown to be effective. Notably, recent work \citep{zou2022large, luo2023model} has shown that DLA \citep{ai2018unbiased} has achieved the state-of-the-art performance on the real-world dataset. DLA jointly learns a propensity model and ranking model to alleviate position bias that can dynamically learn the examination propensities of each position without the need for result randomization \citep{joachims2017unbiased}. Thus, we first adopt DLA to eliminate the position bias. The loss function we utilize is as follows:
\begin{equation}\label{rank_loss}
L_{ranking}= - \sum\limits_{x\in \pi_q, c_q^x=1} \frac{P(o_q^1=1 | \pi_q)}{P(o_q^x=1 | \pi_q)} \cdot \log \frac{e^{f(x)}}{\sum\nolimits_{z\in \pi_q} e^{f(z)}}\text{,}
\end{equation}
\begin{equation}\label{pro_loss}
L_{observation}= - \sum\limits_{x\in \pi_q, c_q^x=1} \frac{P(r_q^1=1 | \pi_q)}{P(r_q^x=1 | \pi_q)} \cdot \log \frac{e^{g(x)}}{\sum\nolimits_{z\in \pi_q} e^{g(z)}}\text{,}
\end{equation}
where $f(x)$ and $g(x)$ denote the output of the ranking model and propensity model respectively and the superscript 1 represents the item in the first place. The probabilities of observation and relevance are computed as follows:
\begin{equation}
    P(r_q^x=1|\pi_q)=\frac{e^{f(x)}}{\sum\nolimits_{z\in \pi_q}e^{f(z)}}\text{,}
\end{equation}
\begin{equation}
    P(o_q^x=1|\pi_q)=\frac{e^{g(x)}}{\sum\nolimits_{z\in \pi_q}e^{g(z)}}\text{.}
\end{equation}
\subsection{Alleviating Selection Bias with Label Correction}
Although position bias can be alleviated by ULTR methods like DLA, it is important to acknowledge that other biases also exist in the click data. Specifically, position bias only accounts for clicked items as non-clicked items are all labeled as 0. However, not all non-clicked items are irrelevant. The documents are not clicked probably because users did not examine them or they are similar to the clicked items. This issue is especially severe when the ranker is effective and most results on the first search result page (SERP) are relevant. Labeling all non-clicked items as 0 could confuse the model in relevance modeling. So, it would be advantageous to adjust the documents labeled as 0 to reasonable non-zero values.

Therefore, we utilize the model trained with DLA as an auxiliary model to generate labels for non-clicked items. We attempt various reasonable strategies to transform the output of this auxiliary model into new labels as shown in Equation(\ref{label_change}). We merely list two representative modes: directly regard the output after the sigmoid function as their labels or set the labels of those whose output is larger than the minimum output of clicked items as one. Here, $l_{ij}, a_{ij}$ means the label, auxiliary model output of the $i^{th}$ query's $j^{th}$ item with $c_{ij}=0$,
\begin{equation}\label{label_change}
l_{ij}=\left\{
\begin{array}{cc}
sig: sigmoid(a_{ij})\\ \\
min: 1 if a_{ij} \geq min(a_{ik}), c_{ik}=1
\end{array}
\right.\text{.}
\end{equation}
\subsection{Negative Sampling}
\label{negative}
Considering a large proportion of non-click items are false negatives, we try to rebuild the ranking lists to reduce the number of false negatives by adding random and hard negatives and replacing original non-click candidate items with random negatives. In this approach, we do not consider position bias and instead treat the original clicked item as relevant, while considering the newly generated negatives as irrelevant.

With the help of Galago \footnote{\url{https://lemur.sourceforge.io/documentation/Galago\%20Documentation.html}}, we pre-generate substantial hard negatives for each query based on BM25 scores. To simulate a multi-level irrelevance among candidate items, these negatives are then sampled from a Gaussian distribution assuming that the BM25 scores follow this distribution. Besides the most relevant (with the largest score) and irrelevant (with the smallest score) hard negatives are also included. We adjust the number of hard negatives from 50 to 90. 

We have devised two schemes: the \textbf{"click-only"} scheme replaces all non-clicked items with random negatives, while the \textbf{"last-click"} scheme replaces items after the last clicked item with random negatives. Moreover, we employ random negatives to ensure that the candidate list for each query is of equal length. The loss function we use is as follows. The $\pi_q^{\prime}$ indicates reconstructed ranking lists to a query $q$,
\begin{equation}
L_{click-only}/L_{last-click}=-\sum_{x\in \pi_q^{\prime}} log\frac{e^{f(x)}}{\sum_{z\in \pi_q^{\prime}}e^{f(z)}}\text{.}
\end{equation}
\subsection{GBDT Training}
In addition to the above methods, we also train a Gradient Boosting Decision Tree (GBDT) model using the validation set with the help of Light Gradient Boosting Machine (LightGBM) \citep{ke2017lightgbm} with LambdaRank objective. Concretely, 80\% of the total validation set for training and 20\% left for validating, split by query id column. By training directly with human annotation labels, which are unbiased, we aim to establish an upper bound for model performance given the same input features. Moreover, the scores obtained from the unbiased learning model can be aggregated with other signals in LightGBM. Consequently, we build two models using two settings of input features, with one including the best model score obtained from previous training with clicks.
\begin{table*}
  \caption{The model performance on both the hidden test set and the validation set. The total validation set (referred to as total valid) was used to evaluate the performance of neural network models, while 20\% of the validation set (referred to as 20\% valid) was used to assess the performance of GBDT models, which were trained using 80\% of the validation set.}
  \label{submitted}
  \begin{tabular}{ccccc}
    \toprule
    Model & & nDCG@10 & & DCG@10\\
   \cline{2-4}
     & test & total valid & 20\% valid & test\\
    \midrule
    \textbf{Scratch-DLA-LC (sig)} & $\textbf{0.5355}$ & $\textbf{0.5019}$ & / & $11.4538$ \\
    Aux-DLA-LC (sig) & $0.5326$ & $0.5015$ & / & $11.3898$ \\
    Scratch-DLA-LC (min) & unk & $0.4816$ & / & unk \\
    Aux-DLA-LC (min) & unk & $0.4947$ & / & unk \\
    DLA & $0.5247$ & $0.4920$ & / & $11.2031$ \\
    lgbBase & $0.5350$ & /  & $0.5003$ & $11.4794$ \\
    lgbBaseAdd & $0.5333$ & / & $0.5021$ & $11.4616$ \\
  \bottomrule
\end{tabular}
\end{table*}
\section{EXPERIMENTS}
We present a detailed explanation of our experimental setup and then summarize the representative results in this section.
\subsection{Experimental Setup}
\label{setup}
\textbf{Methods for comparison.} We compare the performance of the following methods:
\begin{itemize}
\item \textbf{DLA: }The Dual Learning Algorithm \citep{ai2018unbiased} simultaneously learns an unbiased ranking model and a propensity model.
\item \textbf{Scratch-DLA-LC (sig): } Retrain a DLA model using the modified click labels from the first mode (sig) in Equation\ref{label_change}.
\item \textbf{Aux-DLA-LC (sig): }Given a DLA model trained with click labels, continue training this DLA model utilizing the modified click labels from the first mode (sig) in Equation\ref{label_change}.
\item \textbf{Scratch-DLA-LC (min): }Retrain a DLA model using the modified click labels from the second mode (min) in Equation\ref{label_change}.
\item \textbf{Aux-DLA-LC (min): }Given a DLA model trained with click labels, continue training this DLA model utilizing the modified click labels from the second mode (min) in Equation\ref{label_change}.
\item \textbf{lgbBase: }This GBDT model uses human annotation labels, and has the same input features as the unbiased neural ranking models.
\item \textbf{lgbAdd: }Except for the addition of the best model score to the input features, everything is the same as the lgbBase model. 
\end{itemize}
\textbf{Experimental implementation.} According to \citep{mioverview}, the ULTRE-2 organizers provide us with three types of features, traditional word matching features (e.g. TF-IDF, BM25), as well as a score feature and 768 dimensions' embedding features derived from the pre-trained BERT model. We primarily choose the first two types as the input features, i.e. 14 features in total. 

Only results in the initial ranking lists have their associated features. This means that we do not have access to the features of new query-result pairs using the given feature data alone. To address this limitation, we extract another 24 heuristic-based features such as TF, TF-IDF, BM25, and language model with three types of smoothing methods for title, abstract, and the combined fields, the same as features in \citep{yu2023feature} to calculate the features for new query-negative pairs.

When training with click data, for each query, we fix the length of its ranking list as 10. As for negative sampling, at first, the reconstructed ranking list consists of the items we reserve in the initial ranking lists depending on different strategies mentioned in Section \ref{negative} and several random negatives for aligning the length, and on the basis of it, append additional random and hard negatives. We use the total validation set (referred to as total valid) for validating when training the neural network model while for the GBDT model, we use 20\% of the validation set (referred to as 20\% valid) to select the best model. We present the nDCG@10 on both the corresponding validation and test sets.

To implement our models, we utilize the ULTRA framework\footnote{\url{https://github.com/ULTR-Community/ULTRA_pytorch}} as the basis. Our ranking model is a deep neural network (DNN) as well. In our experiment, we start by projecting the above input features to a higher dimension, specifically 64. These projected features are then fed into the DNN with hidden layer dimensions as 32,16 and 8. For optimization, we employ the AdamW optimizer and fine-tune the learning rates in the range of 2e-6 to 1e-5.
\subsection{Experimental Results}
Initially, we evaluate the nDCG@10 of every individual feature mentioned above on the total validation set. Among these features, the best nDCG@10 from given and self-generated traditional features are $0.4608$ and $0.4625$ respectively, and the pre-trained score stands out with an impressive nDCG@10 of $0.4767$, demonstrating its superior significance in model training compared to the traditional features. 

Nevertheless, it's a pity we did not have time to leverage the pre-trained model available on Github\footnote{\url{https://github.com/lixsh6/Tencent_wsdm_cup2023/tree/main/pytorch_unbias}} to obtain the pre-trained score for each query-negative pair. As a result, we solely relied on traditional features as input, making it challenging to compare the effectiveness of model training under negative sampling with other models that include the pre-trained score in their input features. Despite this limitation, our experimental results showed an improvement in the model's performance when using traditional features as input. Consequently, we decided to showcase the nDCG@10 on the total valid to demonstrate the approach's effectiveness.

The nDCG@10 results are shown in Table \ref{submitted}. \emph{unk} denotes we did not submit it to the leaderboard. Besides nDCG@10, we include DCG@10 to compare the results with \citep{chen2023multi} as the pre-trained score derived from the pre-trained model in it. In \citep{chen2023multi}, the best score achieved $10.25$, while in \citep{mioverview} the model without debiasing achieved $11.26$. This verifies the necessity of introducing some proper high-quality traditional features into the model.
\begin{figure}
\centering
\subfigure{
\label{both}
\includegraphics[width=4.0cm,height = 3.0cm]{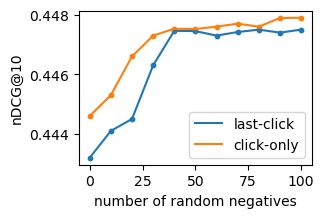}}\subfigure{
\label{click-only}
\includegraphics[width=4.0cm,height = 3.0cm]{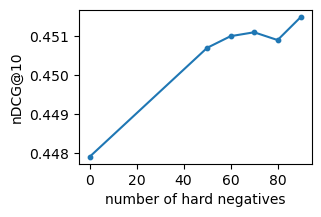}}
\caption{Performance curves of two schemes ("click-only" and "last-click") w.r.t. the number of random and hard negatives. (a) Performance curves of two schemes w.r.t. the number of random negatives. (b) The Performance curve of the "click-only" scheme w.r.t. the number of hard negatives.}
\label{validation}
\end{figure}
\subsubsection{Results on Label Correction and GBDT Training}
From Table \ref{submitted}, we can draw the following conclusions: (1) using already trained DLA as the auxiliary model, combined with both variants can significantly enhance the performance of the model trained from this auxiliary model. And between the two variants, the sigmoid function yields better results, (2) when comparing models with the same input features, the lgbBase model still outperforms the simple DLA model trained with clicks. However, it is not as effective as our proposed Scratch-DLA-LC (sig) model, (3) the performance of the neural network models on the validation set appears to be consistent with their performance on the test set. Thus we could determine the best model based on nDCG@10 on the validation. Nonetheless, there exists inconsistency in the performance of the GBDT model, suggesting potential overfitting issues.
\subsubsection{Results on Negative Sampling}
Results of the negative sampling technique on the total validation set are shown in Figure \ref{strategy}. Comparing the two strategies mentioned in Section \ref{negative}, namely the click-only strategy and the last-click strategy, it is observed that the click-only strategy performs slightly better. Therefore, we choose the click-only strategy as the basis strategy and focus on increasing the number of hard negative samples. From Figure \ref{click-only}, we can observe that by adjusting the number of negative samples, we find that including both random and hard negatives is beneficial, although the improvement is gradual.
\section{CONCLUSIONS}
In this paper, we present our solution for the NTCIR-17 ULTRE-2 task, focusing on the false negative issue. We propose two approaches to tackle this issue. Firstly, we explore the aspect of selection bias and introduce several methods of label correction. Through experiments, we find that our approach, DLA-LC, outperforms the basic DLA model. Additionally, we investigate the use of negative sampling to reconstruct the given ranking lists and reduce the number of false negatives. The nDCG@10 on the validation set indicates that this approach is effective in improving performance. Although we were unable to leverage the important pre-trained score in this experiment, we plan to explore this further in future research to determine the extent of improvement it can bring.
\bibliographystyle{ACM-Reference-Format}
\bibliography{NTCIR-17}

\end{document}